\documentclass{emulateapj}

\shorttitle{CR driven winds}
\shortauthors{Booth et al.}

\newcommand{\msun}{{\rm M}_\odot}

\begin{document}

\title{Simulations of disk galaxies with cosmic ray driven galactic winds}

\author{C. M. Booth\altaffilmark{1}}
\author{Oscar Agertz\altaffilmark{2,1}}
\author{Andrey V. Kravtsov\altaffilmark{1,3,4}}
\author{Nickolay Y. Gnedin\altaffilmark{5,1,3}}
\altaffiltext{1}{Department of Astronomy \& Astrophysics, The University of Chicago, Chicago, IL 60637 USA {\tt cmbooth@oddjob.uchicago.edu}}
\altaffiltext{2}{Department of Physics, University of Surrey, Guildford, GU2 7XH, United Kingdom}
\altaffiltext{3}{Kavli Institute for Cosmological Physics, The University of Chicago, Chicago, IL 60637 USA}
\altaffiltext{4}{Enrico Fermi Institute, The University of Chicago, Chicago, IL 60637 USA}
\altaffiltext{5}{Particle Astrophysics Center, Fermi National Accelerator Laboratory, Batavia, IL 60510 USA}

\begin{abstract}
We present results from high-resolution hydrodynamic simulations of
isolated SMC- and Milky Way-sized galaxies that include a model for
feedback from galactic cosmic rays (CRs).  We find that CRs are
naturally able to drive winds with mass loading factors of up to $\sim
10$ in dwarf systems. The scaling of the mass loading factor with
circular velocity between the two simulated systems is consistent with
$\eta\propto v_{\rm circ}^{1-2}$ required to reproduce the faint end
of the galaxy luminosity function. In addition, simulations with CR
feedback reproduce both the normalization and the slope of the
observed trend of wind velocity with galaxy circular velocity. We find
that winds in simulations with CR feedback exhibit qualitatively
different properties compared to SN driven winds, where most of
acceleration happens violently in situ near star forming sites. The
CR-driven winds are accelerated gently by the large-scale pressure
gradient established by CRs diffusing from the star-forming galaxy
disk out into the halo.  The CR-driven winds also exhibit much cooler
temperatures and, in the SMC-sized system, warm ($T\sim 10^4$~K) gas
dominates the outflow. The prevalence of warm gas in such outflows may
provide a clue as to the origin of ubiquitous warm gas in the gaseous
halos of galaxies detected via absorption lines in quasar spectra.
\end{abstract}


\section{Introduction}

Galactic winds are observed to be ubiquitous in galaxies that have
recently experienced significant amounts star formation \citep[see
  e.g.,][for a review]{veilleux_etal05}.  These outflows represent a
fundamental part of galaxy formation models, because the absence of
outflows galaxy star formation rates (SFRs) are much higher than those
observed \citep[e.g.,][]{stinson_etal13} and baryon fractions in the
disk are close to the universal value \citep[e.g.,][]{crai07}, much
higher than inferred from observations.  In contrast, models that
include a variety of feedback effects predict much lower SFRs and
baryon fractions.  Additionally, outflows are required to drive
metal-enriched gas out of galaxies, as suggested by both observational
\citep[e.g.][]{stei10} and theoretical \citep[e.g.][]{boot12} work.

However, despite their key role in galaxy formation, the exact
processes driving winds remain an open question.  Plausible driving
mechanisms include core collapse supernovae \citep[SN,][]{mart99} and
radiation pressure \citep{murr05,hopk12,ager13}. SN-driven winds are
now routinely included in semi-analytic and numerical simulations.
However, it has long been known that in the disk of the Galaxy there
is a rough equipartition of the magnetic and cosmic ray (CR) energy
densities \citep[e.g][]{beck_krause05}.  This indicates that CRs play
a significant role in dynamics of interstellar medium. Only relatively
recently have the effects of CRs have been considered in the context
of galaxy formation
\citep[e.g.][]{jube08,uhli12,wadepuhl_springel11,sale13} and galaxy
cluster \citep{ensslin_etal07,sijacki_etal08,guo_oh08}
simulations.

A tight link between CRs and star formation is evidenced by the
correlation between a galaxy's infra-red luminosity, closely related
to its SFR, and the luminosity of its radio halo
\citep[e.g.][]{helo85,lack10}.  The relationship is almost linear, has
very little scatter, and does not evolve with redshift \citep{mao11},
indicating that the coupling between star formation and CRs is robust
over a wide variety of conditions.

Although the energy injection rate of CRs is small when compared to
the other sources of energy from star formation, the rate at which
they inject momentum is not \citep{socr08}.  This is because the CRs
that supply most of the pressure in the galaxy generate Alfv\'en waves
in the ISM \citep{went68}, which then scatter the CRs with a mean free
path of $\sim1$\,pc.  Thus, CRs are \lq self-confined\rq\,
\citep[e.g.][]{cesa80}, and it takes $\approx 250$\,Myr for a typical
CR to escape its parent galaxy.  Theoretical models of dynamical
haloes in which CRs diffuse and are advected out in a galactic wind
predict steady, supersonic galaxy-scale outflows driven by a
combination of CR and thermal pressure \citep{brei91,ever08}.

In this Letter we present high-resolution hydrodynamical simulations of
isolated disk galaxies, including a model for the injection, transport
and decay of CRs, to investigate how outflows are driven by CRs
and the properties of the outflowing gas.  

\section{Method}\label{sec:method}

Our simulations are performed with the adaptive-mesh-refinement (AMR)
code {\sc ramses}, described in \citet{teys02}.  The detailed
description of physical processes included in our simulations -- star
formation, radiative cooling, and metal enrichment from Type Ia SNe,
Type II SNe and intermediate mass stars -- can be found in
\citet{ager13}.  SN feedback is modelled by injecting a total of
$10^{51}$\,ergs of thermal energy per SN into the cells neighbouring
the star particle. We do not employ any delay of dissipation for the
injected energy in these runs \citep[the runs are equivalent to the
  ``Energy only'' run in][]{ager13}.

A full description of the CR field would require modelling the
distribution function of CRs as a function of position, momentum and
time.  However, if the CR mean free path is shorter than the length
scale of the problem, the CR field can be described as a fluid
\citep{skil75}.  We thus take the approach of modelling the CR energy
density, $E_{\rm CR}$, as an additional energy field that advects
passively with the gas density \citep[e.g.][]{jone90} and exerts a
pressure $p_{\rm CR}=(\gamma_{\rm CR}-1)E_{\rm CR}$. Thus, the total
pressure entering the momentum and energy equations governing gas
evolution is $p_{\rm tot}=p_{\rm gas}+p_{\rm CR}$.  We assume
throughout that the CR fluid is an ultra relatavistic ideal gas with
$\gamma_{\rm CR}=4/3$.  As described above, CRs undergo a random walk
through the ISM after their injection. Their evolution is thus a
combination of advection with the ambient gas flow and diffusion,
which we parametrize by the diffusion coefficient,
$\kappa=3\times10^{27}\,{\rm cm^2/s}$.

The evolution of baryon and CR fluids is thus governed by the standard
continuity and momentum equations and the following energy equations:

\begin{eqnarray}
\frac{\partial e_{\rm gas}}{\partial t} + \nabla\cdot (e_{\rm gas}v_{\rm gas})& =& -p_{\rm gas}\nabla\cdot v_{\rm gas}+\Gamma - \Lambda_{\rm rad} + \nonumber\\
&&(1-\xi_{\rm CR})\Delta e_{\rm SN}\,,
\end{eqnarray}
\begin{eqnarray}
\frac{\partial e_{\rm CR}}{\partial t} + \nabla\cdot (e_{\rm CR}v_{\rm gas}) &=& -p_{\rm CR}\nabla\cdot v_{\rm gas} + \nabla\cdot (\kappa\cdot \nabla E_{\rm cr})\nonumber\\
&&-\Lambda_{\rm CRcool}+\xi_{\rm CR}\Delta e_{\rm SN}\,,
\end{eqnarray}
where $v_{\rm gas}$ is gas velocity, $p_{\rm gas}$, $e_{\rm gas}$ and
$p_{\rm CR}$, $e_{\rm CR}$ are the pressure and internal energy of gas
and CRs, respectively. The $\Delta e_{\rm SN}$ indicates energy
injection by SN, and $\xi_{\rm CR}$ is the fraction of this energy
that is injected in the form of CRs. $\Lambda_{\rm rad}$ indicates
radiative cooling of gas, while $\Gamma$ indicates the heating of gas
by both CRs and UV radiation. Finally, $\Lambda_{\rm CRcool}$
corresponds to energy losses by CRs both due to decays and Coulomb
interactions with gas mediated by magnetic fields
\citep[e.g.][]{volk96,enss97}.  Following \citet{guo08} we assume that
the CR cooling rate is:
\begin{equation}
\label{eq:crcool}
\Lambda_{\rm CRcool}=-7.51\times10^{-16}\,n_{\rm e}e_{\rm CR}\,{\rm erg}\,{\rm s}^{-1}{\rm cm}^{-3},
\end{equation}
where $n_{\rm e}$ is the local electron number density.  The ratio of
the catastrophic cooling rate to the Couloumb cooling rate for our CR
population is 3.55. Some fraction of the energy lost by the CR
population heats the thermal gas \citep[e.g.][]{mann94} at a rate given by \citep{guo08}
\begin{equation}
\label{eq:crheat}
\Gamma_{{\rm CR Heat}}=2.63\times10^{-16}\,\,n_{\rm e}e_{\rm CR}\,{\rm erg}\,{\rm s}^{-1}{\rm cm}^{-3}.
\end{equation}
Equations \ref{eq:crcool} and \ref{eq:crheat} are solved on every
timestep to calculate the rate of decay of the CR energy density along
with the corresponding gain in the gas thermal energy.

We have tested our CR implementation using a standard shock-tube test
involving gas and CR fluids \citep{pfrommer_etal06} and found that
results accurately match the analytic solution.  Results of this and
other tests will be presented in a forthcoming paper.

Strong shock waves associated with SN explosions have long been
recognized as a likely source of Galactic CRs \citep[e.g.][]{baad34}.
Empirically, in order to match the galactic energy density in CRs, SNe
must be capable of transferring a fraction $\Delta e_{\rm SN}\sim10\%$
of the explosion kinetic energy into the form of CR energy
\citep{hill05}.  In our models we make the assumption that a
certain fraction, $\xi_{\rm CR}=0.1$, of the SN energy is injected to
the CR fluid energy density.  The remaining fraction $1-\xi_{\rm CR}$
is injected thermally into the gas field.

We note that the assumptions that the diffusion of CRs is isotropic
and that the diffusion coefficient is a constant are necessary
simplification in our models, which track neither the direction nor
the strength of the magnetic field.  On small scales ($\sim$100 pc),
the strength of the random component of the galactic magnetic field is
several times higher than the average field strength
\citep[e.g.][]{jans12} because galaxy formation processes
(e.g. supernovae and hydrodynamical turbulence) in the disk
\citep{brei91} and the turbulent dynamo effect and CR buoyancy in the
halo \citep{brei93} tangle the magnetic field to the extent that
isotropic diffusion is a good approximation\citep[e.g.][]{stro07}.
Codes that assume isotropic diffusion are able to predict CR-emitted
spectral data down to the few percent level \citep[e.g][]{orla13}. For
the purposes of this exploratory work we employ the isotropic
diffusion model, but note that investigation of complex models
represents an interesting future direction for this work.

\subsection{The simulation set}

We simulate isolated, model galaxies of two different masses
representing an SMC-sized dwarf galaxy and MW-sized disk galaxy with
three different feedback models: no feedback, thermal feedback only,
and thermal plus CR feedback. The \lq thermal feedback\rq\, runs
inject $100\%$ of the energy released by each SN blast into the gas
thermal energy.  The \lq CR feedback\rq\, runs inject $90\%$ of the SN
energy into the gas thermal energy and the remaining $10\%$ into the
CR energy density field.  Every simulation models radiative cooling,
star-formation and metal enrichment.  All runs are evolved for
0.5\,Gyr and throughout this Letter we report results for this time.

Following \citet{hern93} and \citet{spri00} the galaxy model consists
of a dark matter halo, a stellar bulge and an exponential disc of
stars and gas.  The dark matter halo is modelled as an NFW halo
\citep{nava97}.  The gas and stars are then initialized into an
exponential disk, and the bulge is assumed to have a \citet{hern90}
profile with a scale length that is 10\% of the disk scale length.
The relevant parameters for each set of initial conditions are given
in Table~\ref{tab:ics}.  Each simulation is run with a maximum spatial
resolution of 75\,pc (37.5\,pc) for the MW (SMC) runs.

\begin{table*}
\caption{Parameters of the galaxy models.}                                 
\centering
\begin{tabular}{r l r r r r l l l r r}
 & \multicolumn{4}{c}{Halo Properties} & \multicolumn{6}{|c}{Disk Properties} \\
\hline
Identifier$^{(1)}$ & $m_{\rm 200}^{(2)}$ & $v_{\rm 200}^{(3)}$  & c$^{(4)}$ & $\lambda^{(5)}$ & $f_{\rm g}^{(6)}$ & $M_{\rm gas, disk}^{(7)}$ & $M_{\rm star, disk}^{(8)}$ & $M_{\rm star, bulge}^{(9)}$ & $r_{\rm d}^{(10)}$ & $h_{\rm d}^{(11)}$\\      
           & ($\msun$)  & (${\rm km/s}$) &   &           &            & ($\msun$)  & ($\msun$)  & ($\msun$)   & kpc         & kpc \\
\hline
MW   & $1.1\times10^{12}$ & 150.0 & 10 & 0.02  & 0.20 & $9.0\times10^{9}$ & $3.3\times10^{10}$ & $3.3\times10^{9}$  & 3.6  & 0.36\\        
SMC  & $2.0\times10^9$   & 40.0  & 15 & 0.04  & 0.75 & $4.0\times10^{8}$  & $4.0\times10^{8}$ & $1.0\times10^{7}$  & 0.9  & 0.2 \\        
\end{tabular}\\[1mm]
Notes: From left to right the columns contain: (1) Simulation set name; (2) Spherical overdensity DM halo mass defined relative to the 200 times the critical density at $z=0$; (3) Circular velocity at the virial radius; (4) Concentration of NFW halo; (5) Halo spin parameter; (6) Disk gas fraction; (7) Mass of gas in the disk; (8) Mass of stars in the disk; (9) Mass of stars in the bulge; (10) Scale length of exponential disk; (11) Scale height of gas disk.\\[10mm]
\label{tab:ics}                                                                
\end{table*}                                                                     

\section{Results}\label{sec:results}

\begin{figure}
\center
\includegraphics[width=.98\linewidth,angle=0]{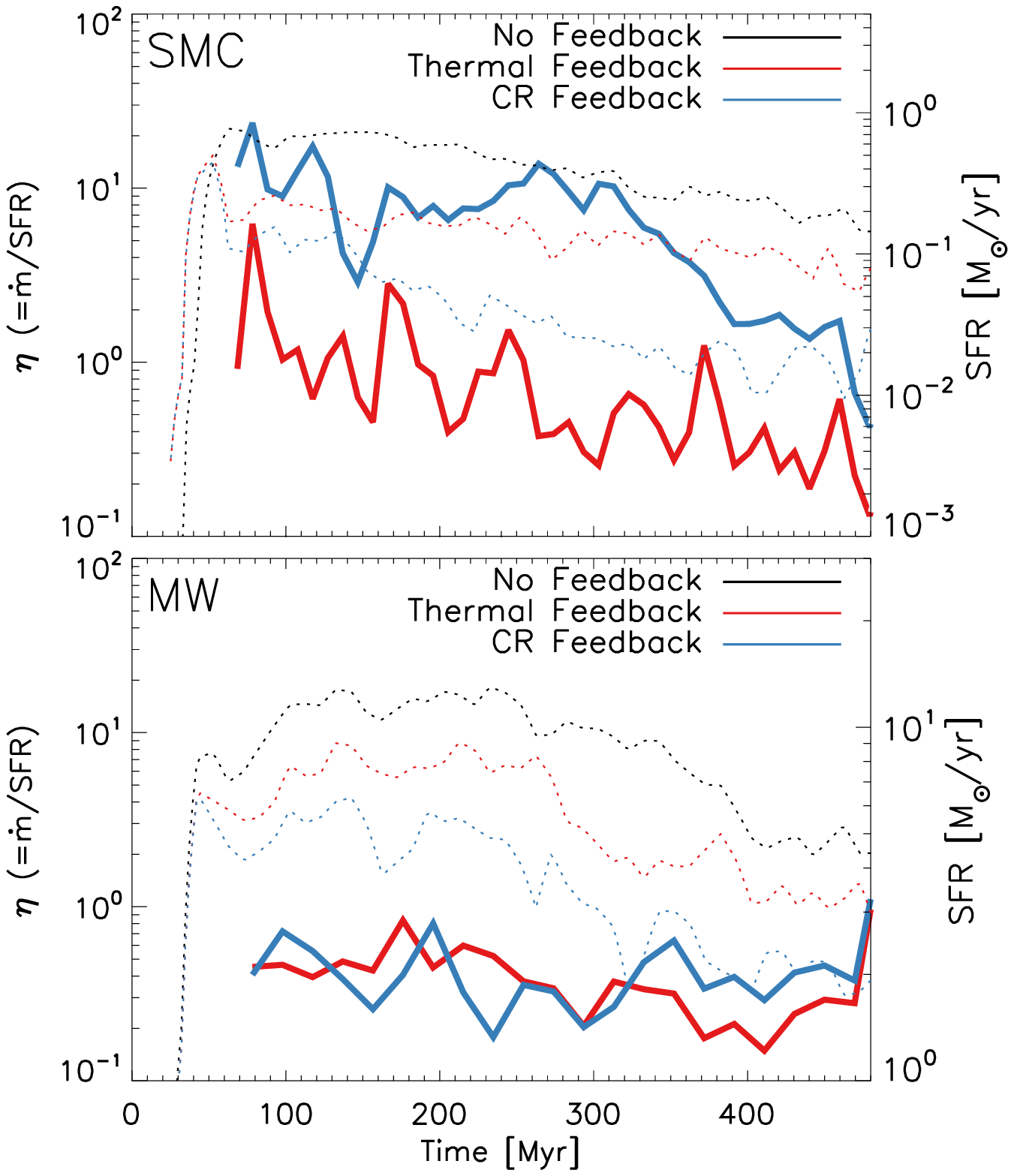}
\caption{\label{fig:ml}The solid curves show the mass loading factor, $\eta$, of the galactic wind, defined as the ratio of the SFR to the gas outflow rate, as a function of time (left-hand axis). The dotted curves show the galaxy SFR (right-hand axis).  The color of each curve denotes the feedback model and the top (bottom) panel shows results for the SMC (MW) simulation.  The no-feedback model (black curves) is not shown on the mass-loading plot because there is a net inflow of gas at all times.  Both feedback models predict mass loadings of $\sim0.5$ for the MW galaxy, but the CR feedback is capable of suppressing the SFR by a larger fraction than the thermal feedback model.  In the SMC galaxy the CR feedback model is capable of driving galactic winds with large ($\sim10$) mass loadings and suppresses the SFR significantly more than thermal feedback alone.\\[1mm]}
\end{figure}

We begin by considering the SFRs of the simulated galaxies in
Fig.~\ref{fig:ml}. The SFR in simulations without feedback is higher
than in simulations with feedback and is higher than typically
observed SFRs of galaxies of these sizes. Simulations with CRs
suppress SFR compared to simulations with thermal SN feedback only,
especially in the SMC-sized galaxy. This is because CRs act as a
source of pressure in the galaxy disk.  This significantly changes the
density PDF of the gas in the disk reducing the fraction of mass in
star forming regions.

Outflow efficiency can be parametrized by the mass loading factor,
$\eta$, defined as the ratio of the gas outflow rate to the SFR.  The
solid curves in Fig.~\ref{fig:ml} show $\eta$ as a function of time
for different simulations.  Outflow rates are measured as the
instantaneous mass flux through the plane parallel to the galactic
disk at a height of 20\,kpc.  In the MW simulation the mass loading is
approximately 0.5 in both simulations, whereas in the SMC simulation
the mass loading is $\sim 10$ in the simulation with CRs and only
$\sim 1$ in the simulation with thermal feedback only.  This indicates
that CRs greatly enhance efficiency of outflows from dwarf galaxies.

\begin{figure}
\center
\includegraphics[width=.98\linewidth,angle=0]{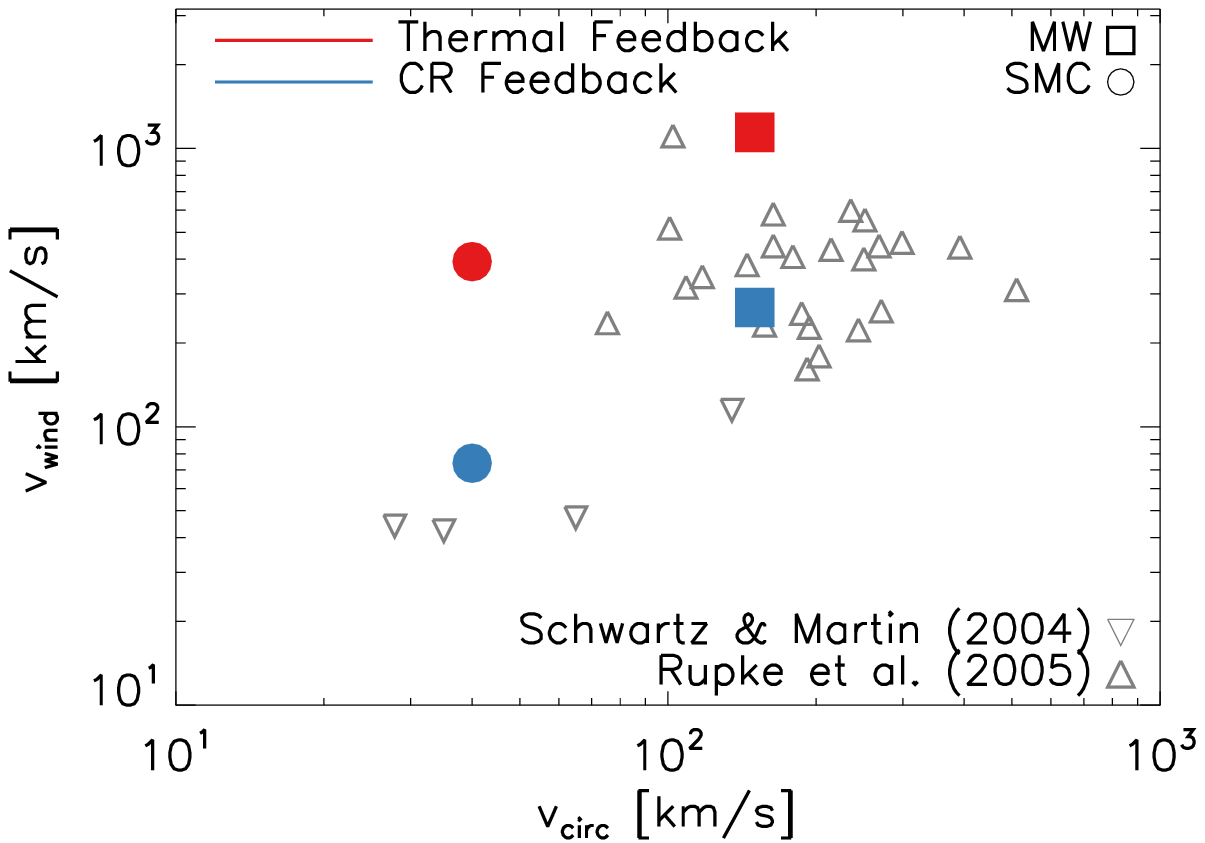}
\caption{\label{fig:vz}Velocity of the outflowing gas ($w_{\rm wind}$) as a function of halo circular velocity.  The gray points show the observations of \citet{schw04} (downward pointing triangles) and \citet{rupk05} (upward pointing triangles).  The solid points show simulation predictions.  The squares (circles) show the MW (SMC) simulations and the colors denote the feedback model.  In both galaxies, the outflows in the CR feedback models (blue points) have velocities comparable to the obserations, whereas the thermal feedback models (red points) overestimate the wind velocity by a large factor.\\[1mm]}
\end{figure}

Figure~\ref{fig:vz} shows velocity of the outflowing gas, $v_{\rm
  wind}$, as a function of the circular velocity of the halo, $v_{\rm
  circ}$, compared to observations of cool wind gas around dwarf
galaxies \citep{schw04} and $z<0.5$ starburst dominated galaxies
\citep{rupk05}.  We measure outflow velocities by projecting the gas
field perpendicular to the disk and calculating the velocity that
contains 90\% of the cool ($T<10^5$\,K) gas.  In each galaxy the
thermal feedback simulation predicts outflow velocities that are
significantly larger than those observed whereas the CR runs are
comparable to the observations.

\begin{figure*}
\center
\includegraphics[width=.98\linewidth,angle=0]{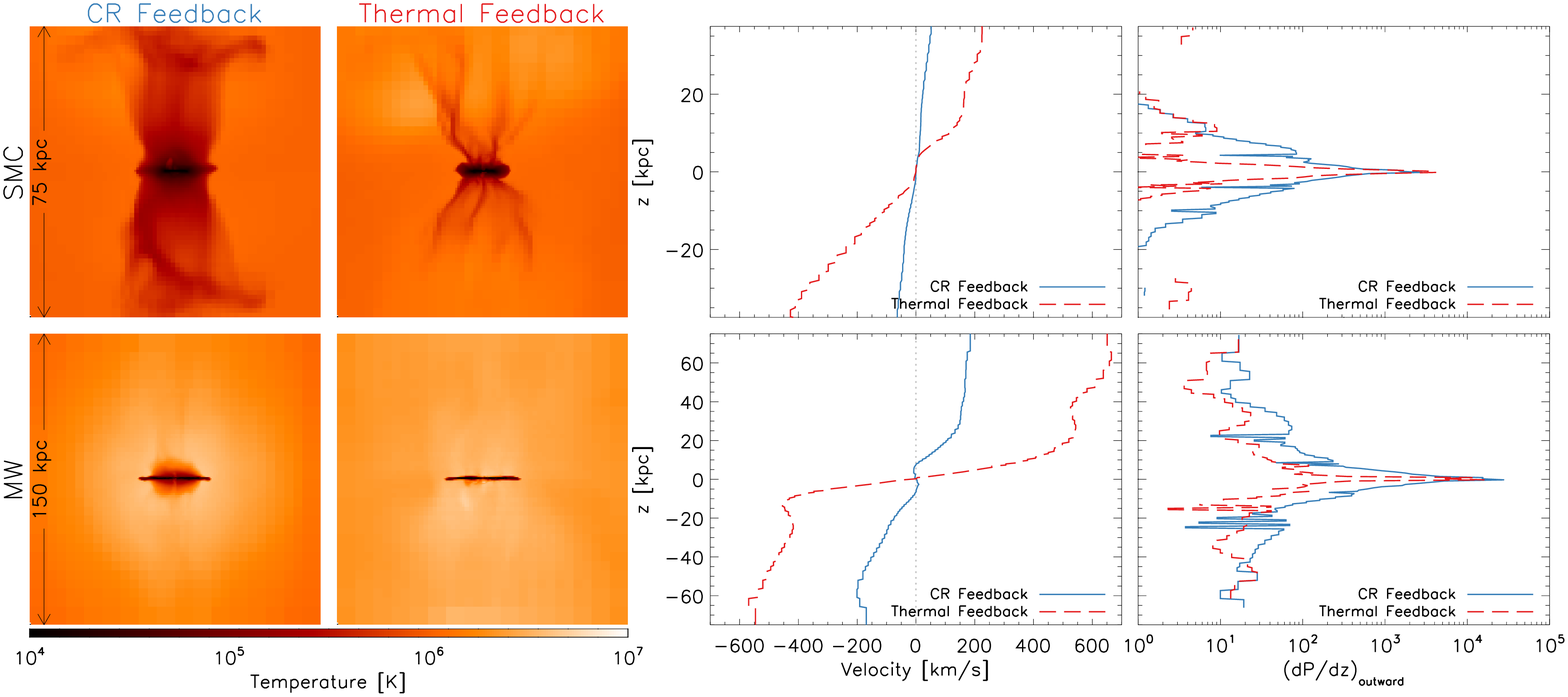}
\caption{\label{fig:im}Edge-on maps of the temperature in a thin slice around the MW (\emph{top panels}) and SMC galaxies (\emph{bottom panels}) for both the thermal feedback \emph{left panels}) and CR feedback (\emph{right panels}).  CR feedback has a large effect on the temperature structure of the halo gas.  The plots show the median velocity (\emph{left panels}) and outward pressure force (\emph{right panels}) as a function of height from the disk for the same two simulations.  All quantities are calculated in a cylinder of radius 3kpc, centered on the galactic disk.  It is clear that the effect of the CRs is to increase the outward pressure forces in the halo by a factor of 3-5 at all z.  This pressure gradient slowly accelerates the wind into the halo.  The wind in the thermal feedback simulations is accelerated abruptly from the disk and maintains a constant velocity thereafter.}
\end{figure*}

Finally, Fig.~\ref{fig:im} shows the temperature of the outflowing gas
in a thin slice through the centre of the simulated galaxies (left).
The notable difference between simulations is that wind in the CR
simulation is considerably cooler, especially in the SMC simulation.
The panels to the right of this figure show the profiles of velocity
and outward pressure gradient.  The thermal feedback run has winds
that accelerate abruptly from the galactic disk up to $\sim 700$\,km/s
and thereafter have a constant velocity.  The CR simulations, however
show a wind that accelerates smoothly into the halo.  The reason for
this is revealed in the right-hand panels, where it is immediately
apparent that the pressure gradient in the halo with CRs is a factor
of 3-10 larger in the CR simulation than in the thermal feedback
simulation (the difference is particularly striking in the SMC
simulation). These results illustrate that the wind properties in the
simulations with CRs are qualitatively different properties to the
wind driven by thermal SN feedback.

\section{Discussion and conclusions}\label{sec:conclusions}

Our simulations show that energy injection in the form of CRs is a
promising feedback process that can substantially aid in driving
outflows from star-forming galaxies. First, we find that CR injection
can suppress the SFR by providing an extra source of pressure that
stabilizes the disk. Turbulent and CR pressure are in equipartition in
the disk, thus the CR pressure can significantly affect most of the
volume of the disk, but will be sub-dominant inside supersonic
molecular clouds, where turbulent pressure dominates over both CR and
thermal presure, particularly in the dwarf galaxy. The SFRs measured
in our galaxies with CR feedback are comparable to observed SFRs for
both the MW and the SMC.

Second, we find that addition of the CR feedback increases the mass
loading factor, $\eta$, in the dwarf galaxy by a factor of ten
compared to the simulation with SN only feedback. As a result, the SMC
and MW-sized galaxies (circular velocities of $40$ and $150$ km/s,
respectively) have mass loading factors that differ by a factor of
$\sim 3-10$, depending on the stage of evolution. This is in rough
agreement with expectations from theoretical models based on
simulations and semi-analytic models, which show that dependence
$\eta\propto v_{\rm circ}^{\alpha}$ with $\alpha\sim 1-2$ is needed to
reproduce the observed faint end of the galaxy stellar mass function
and other properties of the galaxy population
\citep[e.g.,][]{some08,scha10,oppe10,dutt12}. Moreover, the wind
velocities in the SMC and MW-sized simulated galaxies are consistent
with the observed trend for galaxies in this mass range
\citep{schw04,rupk05} both in normalization and slope. Although we
have reported only two models, these results are encouraging,
especially because simulation parameters have not been tuned to
reproduce these observations.

Perhaps the most intriguing difference of the CR-driven winds compared
to the winds driven by thermal SN feedback is that they contain
significantly more ``warm'' $T\sim 10^4$~K gas. This is especially
true for the dwarf galaxy, which develops a wind strikingly colder
than in the SN-only simulation (see Fig.~\ref{fig:im}). The CR-driven
wind has a lower velocity, and is accelerated gradually with vertical
distance from the disk. The reason for these differences is that the
gas ejected from the disk is accelerated not only near star-forming
regions, as is the case in SN-only simulations, but is continuously
accelerated by the pressure gradient established by CRs diffused
outside of the disk (see Fig.~\ref{fig:im}). The diffusion of CRs is
thus a key factor in ejecting winds and in their resulting colder
temperatures.  The cooler temperatures of the ejected gas may be one
of the most intriguing new features of the CR-driven winds, as this
may provide a clue on the origin of ubiquitous warm gas in gaseous
halos of galaxies \citep[e.g.,][and references
  therein]{chen12}. Detailed predictions of CGM properties will
require cosmological galaxy formation simulations incorporating CR
feedback, which we will pursue in future work.

Several studies have explored effects of CR injection on
galaxies. \citet{jube08} found that CRs suppress the SFR in dwarf
galaxies by an amount comparable to our simulations, but have almost
no effect on the SFR of MW-sized systems. We find significant SFR
suppression for both masses. Additionally, \citet{jube08} found that
CRs did not generate winds with diffusion alone and in a recent study
using a similar model \citet{uhli12} argued that to launch winds CR
streaming is crucial. In contrast, we find that CR-driven winds are
established with CR diffusion alone. These differences likely arise
due to assumption of equilibrium between the sources of CRs (star
formation $\propto \rho^{1.5}$) and the sinks (catastrophic losses
$\propto\rho^{-1}$) in the subgrid model of \citet{jube08}. The
subgrid model thus predicts that CR pressure scales as $\sqrt{\rho}$
and is subdominant to the thermal ISM pressure at densities $n_{\rm
  H}>0.2$\,cm$^{-3}$ \citep[see Fig. 7 of][]{jube08}. This assumption
of equilibrium, which is likely true only in the deepest parts of the
galaxy potential well \citep[see e.g. the discussion in][]{socr08},
breaks down in lower density gas.  In our simulations we do not assume
such equilibrium and we find significant pressure contributions from
CRs up to much higher densities. Our results thus indicate that CRs,
even in the diffusion limit, not only suppress star formation but
also drive outflows efficiently. Thus, the effects of CR feedback on
the properties of galaxies of different masses should be significantly
stronger and span a wider range of masses than simulations that use
the \citet{jube08} model \citep[e.g.][]{wadepuhl_springel11}.
  
While this manuscript was in a late stage of preparation
\citet{sale13} appeared as a preprint. These authors have presented
simulations of a MW-sized galaxy, similar to the model presented here,
albeit without accounting for CR cooling losses and with a much larger
SFR in their model galaxy (up to $\sim 200-300\ M_{\odot}\,\rm
yr^{-1}$.  Where our results overlap (e.g., mass-loading factor) with
those of \citet{sale13} we find remarkably good agreement. These
authors also find that outflows are efficiently generated with CR
diffusion alone. Our study extends the results of \citet{sale13} by
presenting the differences between wind properties in dwarf and
MW-sized systems. The results of \citet{sale13} and our study indicate
that CRs can significantly suppress star formation in galaxies and
efficiently drive outflows with significant mass loading factors and
velocities comparable to observed outflows. A detailed exploration of
the effects of such feedback on the galaxy population in a full
cosmological setting is therefore extremely interesting.

\acknowledgements NG and AK were supported via NSF grant OCI-0904482.
AK was supported by NASA ATP grant NNH12ZDA001N and by the Kavli
Institute for Cosmological Physics at the University of Chicago
through grants NSF PHY-0551142 and PHY-1125897 and an endowment from
the Kavli Foundation and its founder Fred Kavli.


\begin{thebibliography}{}
\expandafter\ifx\csname natexlab\endcsname\relax\def\natexlab#1{#1}\fi

\bibitem[{{Agertz} {et~al.}(2013){Agertz}, {Kravtsov}, {Leitner}, \&
  {Gnedin}}]{ager13}
{Agertz}, O., {Kravtsov}, A.~V., {Leitner}, S.~N., \& {Gnedin}, N.~Y. 2013,
  ApJ, 770, 25

\bibitem[{{Baade} \& {Zwicky}(1934)}]{baad34}
{Baade}, W., \& {Zwicky}, F. 1934, Physical Review, 46, 76

\bibitem[{{Beck} \& {Krause}(2005)}]{beck_krause05}
{Beck}, R., \& {Krause}, M. 2005, Astronomische Nachrichten, 326, 414

\bibitem[{{Booth} {et~al.}(2012){Booth}, {Schaye}, {Delgado}, \& {Dalla
  Vecchia}}]{boot12}
{Booth}, C.~M., {Schaye}, J., {Delgado}, J.~D., \& {Dalla Vecchia}, C. 2012,
  MNRAS, 420, 1053

\bibitem[{{Breitschwerdt} {et~al.}(1991){Breitschwerdt}, {McKenzie}, \&
  {Voelk}}]{brei91}
{Breitschwerdt}, D., {McKenzie}, J.~F., \& {Voelk}, H.~J. 1991, A\&A, 245, 79

\bibitem[{{Breitschwerdt} {et~al.}(1993){Breitschwerdt}, {McKenzie}, \&
  {Voelk}}]{brei93}
---. 1993, A\&A, 269, 54

\bibitem[{{Cesarsky}(1980)}]{cesa80}
{Cesarsky}, C.~J. 1980, ARA\&A, 18, 289

\bibitem[{{Chen}(2012)}]{chen12}
{Chen}, H.-W. 2012, \mnras, 427, 1238

\bibitem[{{Crain} {et~al.}(2007){Crain}, {Eke}, {Frenk}, {Jenkins}, {McCarthy},
  {Navarro}, \& {Pearce}}]{crai07}
{Crain}, R.~A., {Eke}, V.~R., {Frenk}, C.~S., {et~al.} 2007, MNRAS, 377, 41

\bibitem[{{Dutton}(2012)}]{dutt12}
{Dutton}, A.~A. 2012, MNRAS, 424, 3123

\bibitem[{{Ensslin} {et~al.}(1997){Ensslin}, {Biermann}, {Kronberg}, \&
  {Wu}}]{enss97}
{Ensslin}, T.~A., {Biermann}, P.~L., {Kronberg}, P.~P., \& {Wu}, X.-P. 1997,
  ApJ, 477, 560

\bibitem[{{En{\ss}lin} {et~al.}(2007){En{\ss}lin}, {Pfrommer}, {Springel}, \&
  {Jubelgas}}]{ensslin_etal07}
{En{\ss}lin}, T.~A., {Pfrommer}, C., {Springel}, V., \& {Jubelgas}, M. 2007,
  \aap, 473, 41

\bibitem[{{Everett} {et~al.}(2008){Everett}, {Zweibel}, {Benjamin}, {McCammon},
  {Rocks}, \& {Gallagher}}]{ever08}
{Everett}, J.~E., {Zweibel}, E.~G., {Benjamin}, R.~A., {et~al.} 2008, ApJ, 674,
  258

\bibitem[{{Guo} \& {Oh}(2008{\natexlab{a}})}]{guo_oh08}
{Guo}, F., \& {Oh}, S.~P. 2008{\natexlab{a}}, \mnras, 384, 251

\bibitem[{{Guo} \& {Oh}(2008{\natexlab{b}})}]{guo08}
---. 2008{\natexlab{b}}, MNRAS, 384, 251

\bibitem[{{Helou} {et~al.}(1985){Helou}, {Soifer}, \&
  {Rowan-Robinson}}]{helo85}
{Helou}, G., {Soifer}, B.~T., \& {Rowan-Robinson}, M. 1985, ApJL, 298, L7

\bibitem[{{Hernquist}(1990)}]{hern90}
{Hernquist}, L. 1990, ApJ, 356, 359

\bibitem[{{Hernquist}(1993)}]{hern93}
---. 1993, ApJS, 86, 389

\bibitem[{{Hillas}(2005)}]{hill05}
{Hillas}, A.~M. 2005, Journal of Physics G Nuclear Physics, 31, 95

\bibitem[{{Hopkins} {et~al.}(2012){Hopkins}, {Quataert}, \& {Murray}}]{hopk12}
{Hopkins}, P.~F., {Quataert}, E., \& {Murray}, N. 2012, MNRAS, 421, 3522

\bibitem[{{Jansson} \& {Farrar}(2012)}]{jans12}
{Jansson}, R., \& {Farrar}, G.~R. 2012, ApJL, 761, L11

\bibitem[{{Jones} \& {Kang}(1990)}]{jone90}
{Jones}, T.~W., \& {Kang}, H. 1990, ApJ, 363, 499

\bibitem[{{Jubelgas} {et~al.}(2008){Jubelgas}, {Springel}, {En{\ss}lin}, \&
  {Pfrommer}}]{jube08}
{Jubelgas}, M., {Springel}, V., {En{\ss}lin}, T., \& {Pfrommer}, C. 2008, A\&A,
  481, 33

\bibitem[{{Lacki} {et~al.}(2010){Lacki}, {Thompson}, \& {Quataert}}]{lack10}
{Lacki}, B.~C., {Thompson}, T.~A., \& {Quataert}, E. 2010, ApJ, 717, 1

\bibitem[{{Mannheim} \& {Schlickeiser}(1994)}]{mann94}
{Mannheim}, K., \& {Schlickeiser}, R. 1994, A\&A, 286, 983

\bibitem[{{Mao} {et~al.}(2011){Mao}, {Huynh}, {Norris}, {Dickinson}, {Frayer},
  {Helou}, \& {Monkiewicz}}]{mao11}
{Mao}, M.~Y., {Huynh}, M.~T., {Norris}, R.~P., {et~al.} 2011, ApJ, 731, 79

\bibitem[{{Martin}(1999)}]{mart99}
{Martin}, C.~L. 1999, ApJ, 513, 156

\bibitem[{{Murray} {et~al.}(2005){Murray}, {Quataert}, \& {Thompson}}]{murr05}
{Murray}, N., {Quataert}, E., \& {Thompson}, T.~A. 2005, ApJ, 618, 569

\bibitem[{{Navarro} {et~al.}(1997){Navarro}, {Frenk}, \& {White}}]{nava97}
{Navarro}, J.~F., {Frenk}, C.~S., \& {White}, S.~D.~M. 1997, ApJ, 490, 493

\bibitem[{{Oppenheimer} {et~al.}(2010){Oppenheimer}, {Dav{\'e}}, {Kere{\v s}},
  {Fardal}, {Katz}, {Kollmeier}, \& {Weinberg}}]{oppe10}
{Oppenheimer}, B.~D., {Dav{\'e}}, R., {Kere{\v s}}, D., {et~al.} 2010, MNRAS,
  406, 2325

\bibitem[{{Orlando} \& {Strong}(2013)}]{orla13}
{Orlando}, E., \& {Strong}, A.~W. 2013, ArXiv e-prints, arXiv:1307.2264

\bibitem[{{Pfrommer} {et~al.}(2006){Pfrommer}, {Springel}, {En{\ss}lin}, \&
  {Jubelgas}}]{pfrommer_etal06}
{Pfrommer}, C., {Springel}, V., {En{\ss}lin}, T.~A., \& {Jubelgas}, M. 2006,
  \mnras, 367, 113

\bibitem[{{Rupke} {et~al.}(2005){Rupke}, {Veilleux}, \& {Sanders}}]{rupk05}
{Rupke}, D.~S., {Veilleux}, S., \& {Sanders}, D.~B. 2005, ApJS, 160, 115

\bibitem[{{Salem} \& {Bryan}(2013)}]{sale13}
{Salem}, M., \& {Bryan}, G.~L. 2013, ArXiv e-prints, arXiv:1307.6215

\bibitem[{{Schaye} {et~al.}(2010){Schaye}, {Dalla Vecchia}, {Booth}, \& {et
  al.}}]{scha10}
{Schaye}, J., {Dalla Vecchia}, C., {Booth}, C.~M., \& {et al.} 2010, MNRAS,
  402, 1536

\bibitem[{{Schwartz} \& {Martin}(2004)}]{schw04}
{Schwartz}, C.~M., \& {Martin}, C.~L. 2004, ApJ, 610, 201

\bibitem[{{Sijacki} {et~al.}(2008){Sijacki}, {Pfrommer}, {Springel}, \&
  {En{\ss}lin}}]{sijacki_etal08}
{Sijacki}, D., {Pfrommer}, C., {Springel}, V., \& {En{\ss}lin}, T.~A. 2008,
  \mnras, 387, 1403

\bibitem[{{Skilling}(1975)}]{skil75}
{Skilling}, J. 1975, MNRAS, 172, 557

\bibitem[{{Socrates} {et~al.}(2008){Socrates}, {Davis}, \&
  {Ramirez-Ruiz}}]{socr08}
{Socrates}, A., {Davis}, S.~W., \& {Ramirez-Ruiz}, E. 2008, ApJ, 687, 202

\bibitem[{{Somerville} {et~al.}(2008){Somerville}, {Hopkins}, {Cox},
  {Robertson}, \& {Hernquist}}]{some08}
{Somerville}, R.~S., {Hopkins}, P.~F., {Cox}, T.~J., {Robertson}, B.~E., \&
  {Hernquist}, L. 2008, MNRAS, 391, 481

\bibitem[{{Springel}(2000)}]{spri00}
{Springel}, V. 2000, MNRAS, 312, 859

\bibitem[{{Steidel} {et~al.}(2010){Steidel}, {Erb}, {Shapley}, {Pettini},
  {Reddy}, {Bogosavljevi{\'c}}, {Rudie}, \& {Rakic}}]{stei10}
{Steidel}, C.~C., {Erb}, D.~K., {Shapley}, A.~E., {et~al.} 2010, ApJ, 717, 289

\bibitem[{{Stinson} {et~al.}(2013){Stinson}, {Brook}, {Macci{\`o}}, {Wadsley},
  {Quinn}, \& {Couchman}}]{stinson_etal13}
{Stinson}, G.~S., {Brook}, C., {Macci{\`o}}, A.~V., {et~al.} 2013, \mnras, 428,
  129

\bibitem[{{Strong} {et~al.}(2007){Strong}, {Moskalenko}, \& {Ptuskin}}]{stro07}
{Strong}, A.~W., {Moskalenko}, I.~V., \& {Ptuskin}, V.~S. 2007, Annual Review
  of Nuclear and Particle Science, 57, 285

\bibitem[{{Teyssier}(2002)}]{teys02}
{Teyssier}, R. 2002, A\&A, 385, 337

\bibitem[{{Uhlig} {et~al.}(2012){Uhlig}, {Pfrommer}, {Sharma}, {Nath},
  {En{\ss}lin}, \& {Springel}}]{uhli12}
{Uhlig}, M., {Pfrommer}, C., {Sharma}, M., {et~al.} 2012, MNRAS, 423, 2374

\bibitem[{{Veilleux} {et~al.}(2005){Veilleux}, {Cecil}, \&
  {Bland-Hawthorn}}]{veilleux_etal05}
{Veilleux}, S., {Cecil}, G., \& {Bland-Hawthorn}, J. 2005, \araa, 43, 769

\bibitem[{{V{\"o}lk} {et~al.}(1996){V{\"o}lk}, {Aharonian}, \&
  {Breitschwerdt}}]{volk96}
{V{\"o}lk}, H.~J., {Aharonian}, F.~A., \& {Breitschwerdt}, D. 1996, Space
  Science Reviews, 75, 279

\bibitem[{{Wadepuhl} \& {Springel}(2011)}]{wadepuhl_springel11}
{Wadepuhl}, M., \& {Springel}, V. 2011, \mnras, 410, 1975

\bibitem[{{Wentzel}(1968)}]{went68}
{Wentzel}, D.~G. 1968, ApJ, 152, 987

\end{thebibliography}
\end{document}